\documentclass[runningheads]{svmult}
\usepackage{makeidx}   
\usepackage{graphicx}  
\usepackage{subeqnar}  
\usepackage{multicol}  
\usepackage{cropmark} 
\usepackage{physprbb}  
\newcommand{\greeksym}[1]{{\usefont{U}{psy}{m}{n}#1}}
\newcommand{\umu}{\mbox{\greeksym{m}}}

\begin{document} 
\title*{Searching for Supersymmetric Dark Matter-
\protect\newline The Directional Rate for Caustic Rings.
}
\titlerunning{ Searching for Supersymmetric Dark Matter}
%
%
\author{J. D. Vergados} 
\authorrunning{J. D. Vergados} 
\institute{ Theoretical Physics Division, University of Ioannina, 
GR-45110, 
Greece\\E-mail:Vergados@cc.uoi.gr} 
\maketitle              
\begin{abstract} 
The detection of the theoretically expected dark matter
\index{Dark Matter Detection} 
is central to particle physics and cosmology. Current fashionable supersymmetric
models provide a natural dark matter candidate which is the lightest
supersymmetric particle (LSP). 
The theoretically obtained event rates
are usually  very low or even undetectable.
 So the experimentalists would like to exploit special signatures
like the directional rates and the modulation effect. 
In the
present paper we study these signatures focusing on a specific class of
 non-isothermal models
involving flows of caustic rings.
\end{abstract}

\section{Introduction}
In recent years the consideration of exotic dark matter has become necessary
in order to close the Universe  \cite{Jungm}.  Recent data from the 
 High-z Supernova Search Team \cite{HSST} and the
Supernova Cosmology Project  ~\cite {SPF}
$^,$~\cite {SCP} suggest the presence of a cosmological connstanta
 $\Lambda$. In fact 
 the situation can be adequately described by 
 a barionic component $\Omega_B=0.1$ along with 
$\Omega _{CDM}= 0.3$ and $\Omega _{\Lambda}= 0.6$ (see also
Turner, these proceedings). 

 Since this particle is expected to be very massive, $m_{\chi} \geq 30 GeV$, and
extremely non relativistic with average kinetic energy $T \leq 100 KeV$,
it can be directly detected ~\cite{JDV}$^-$\cite{KVprd} mainly via the
recoiling nucleus.

 Using an effective supersymmetric Lagrangian at the  
quark level, see e.g. Jungman et al 
~\cite{Jungm} and references therein , a quark model for the nucleon
 ~\cite{Dree,Chen}
and nuclear  wave functions ~\cite{KVprd} one can obtain the needed
detection rates.  They are typically very low. So experimentally one 
would like to exploit the modulation of the event rates due to the 
earth's revolution around the sun. In our previous work
\cite{JDV99}$^,$ \cite{JDV99b} we found
enhanced modulation, if one uses appropriate asymmetric velocity
distribution.  
 The isolated galaxies are, however, surrounded by cold dark matter
, which,  due to gravity, keeps falling continuously on 
them from all directions \cite{SIKIVIE}.
It is the purpose of our present paper to exploit the results of
such a scenario.

\section{The Basic Ingredients for LSP Nucleus Scattering}

The differential cross section can be cast in the form 
\cite{JDV99b}:
\begin{equation}
d\sigma (u,\upsilon)= \frac{du}{2 (\mu _r b\upsilon )^2} [(\bar{\Sigma} _{S} 
                   +\bar{\Sigma} _{V}~ \frac{\upsilon^2}{c^2})~F^2(u)
                       +\bar{\Sigma} _{spin} F_{11}(u)]
\label{2.9}
\end{equation}
\begin{equation}
\bar{\Sigma} _{S} = \sigma_0 (\frac{\mu_r}{m_N})^2  \,
 \{ A^2 \, [ (f^0_S - f^1_S \frac{A-2 Z}{A})^2 \, ] \simeq \sigma^S_{p,\chi^0}
        A^2 (\frac{\mu_r}{m_N})^2 
\label{2.10}
\end{equation}
\begin{equation}
\bar{\Sigma} _{spin}  =  \sigma^{spin}_{p,\chi^0}~\zeta_{spin}
\label{2.10a}
\end{equation}
\begin{equation}
\zeta_{spin}= \frac{(\mu_r/m_N)^2}{3(1+\frac{f^0_A}{f^1_A})^2}
[(\frac{f^0_A}{f^1_A} \Omega_0(0))^2 \frac{F_{00}(u)}{F_{11}(u)}
  +  2\frac{f^0_A}{ f^1_A} \Omega_0(0) \Omega_1(0)
\frac{F_{01}(u)}{F_{11}(u)}+  \Omega_1(0))^2  \, ] 
\label{2.10b}
\end{equation}
\begin{equation}
\bar{\Sigma} _{V}  =  \sigma^V_{p,\chi^0}~\zeta_V 
\label{2.10c}
\end{equation}
\begin{equation}
\zeta_V = \frac{(\mu_r/m_N)^2} {(1+\frac{f^1_V}{f^0_V})^2} A^2 \, 
(1-\frac{f^1_V}{f^0_V}~\frac{A-2 Z}{A})^2 [ (\frac{\upsilon_0} {c})^2  
[ 1  -\frac{1}{(2 \mu _r b)^2} \frac{2\eta +1}{(1+\eta)^2} 
\frac{\langle~2u~ \rangle}{\langle~\upsilon ^2~\rangle}] 
\label{2.10d}
\end{equation}
\\
$\sigma^i_{p,\chi^0}=$ proton cross-section with $i=S,spin,V$ given by:\\
$\sigma^S_{p,\chi^0}= \sigma_0 ~(f^0_S)^2$   (scalar) , 
(the isovector scalar is negligible, i.e. $\sigma_p^S=\sigma_n^S)$\\
$\sigma^{spin}_{p,\chi^0}= \sigma_0~~3~(f^0_A+f^1_A)^2$
  (spin) ,
$\sigma^{V}_{p,\chi^0}= \sigma_0~(f^0_V+f^1_V)^2$  
(vector)   \\
where $m_p$ is the proton mass,
 $\eta = m_x/m_N A$, and
 $\mu_r$ is the reduced mass and  
\begin{equation}
\sigma_0 = \frac{1}{2\pi} (G_F m_N)^2 \simeq 0.77 \times 10^{-38}cm^2 
\label{2.7} 
\end{equation}
\begin{equation}
u = q^2b^2/2~~or~~
Q=Q_{0}u, \qquad Q_{0} = \frac{1}{A m_{N} b^2} 
\label{2.15} 
\end{equation}
where
b is (the harmonic oscillator) size parameter, 
q (Q) is the momentum (energy) transfer to the nucleus.
In the above expressions $F(u)$ is the nuclear form factor and
$F_{\rho \rho^{\prime}}(u)$
are the spin form factors \cite{KVprd} ($\rho , \rho^{'}$ are isospin indices)
The differential non-directional  rate can be written as
\begin{equation}
dR=dR_{non-dir} = \frac{\rho (0)}{m_{\chi}} \frac{m}{A m_N} 
d\sigma (u,\upsilon) | {\boldmath \upsilon}|
\label{2.18}  
\end{equation}
 where
 $\rho (0) = 0.3 GeV/cm^3$ is the LSP density in our vicinity and 
 m is the detector mass 

 The directional differential rate \cite{Copi99} in the direction
$\hat{e}$ is 
given by :
\begin{equation}
dR_{dir} = \frac{\rho (0)}{m_{\chi}} \frac{m}{A m_N} 
{\boldmath \upsilon}.\hat{e} H({\boldmath \upsilon}.\hat{e})
 ~\frac{1}{2 \pi}~  
d\sigma (u,\upsilon)
\label{2.20}  
\end{equation}
where H the Heaviside step function. The factor of $1/2 \pi$ is 
introduced, since we have chosen to normalize our results to the
usual differential rate.
 
 We will now  examine the consequences of the earth's
revolution around the sun (the effect of its rotation around its axis is
expected to be negligible) i.e. the modulation effect. 

 Following Sikivie we will consider $2 \times N$ caustic rings, (i,n)
, i=(+.-) and n=1,2,...N (N=20 in the model of Sikivie et al),
each of which
contributes to the local density a fraction $\bar{\rho}_n$ of the
the summed density $\bar{\rho}$ of each type $i=+,-$. and has
velocity ${\bf y}_n=(y_{nx},y_{ny},y_{nz})$
, in units of $\upsilon_0=220~Km/s$, with respect to the
galactic center.

We find it convenient to choose the z-axis 
in the direction of the motion of the
the sun, the y-axis is normal to the plane of the galaxy and 
the x-axis is in the radial direction. 
The needed quantities are taken from the 
work of Sikivie (table 1 of last Ref. \cite{SIKIVIE}) by the 
definitions
$y_n=\upsilon_n/\upsilon_0
,y_{nz}=\upsilon_{n\phi}/\upsilon_0
,y_{nx}=\upsilon_{nr}/\upsilon_0
,y_{ny}=\upsilon_{nz}/\upsilon_0$
. This leads to a
velocity distribution of the form:
\begin{equation}
f(\upsilon^{\prime}) = \sum_{n=1}^N~\delta({\bf \upsilon} ^{'}
    -\upsilon_0~{\bf y}_n)
\label{3.1} 
\end{equation}
The velocity of the earth around the
sun is given by \cite{KVprd}. 
\begin{equation}
{\bf \upsilon}_E \, = \, {\bf \upsilon}_0 \, + \, {\bf \upsilon}_1 \, 
= \, {\bf \upsilon}_0 + \upsilon_1(\, sin{\alpha} \, {\bf \hat x}
-cos {\alpha} \, cos{\gamma} \, {\bf \hat y}
+cos {\alpha} \, sin{\gamma} \, {\bf \hat z} \,)
\label{3.6}  
\end{equation}
where $\alpha$ is the phase of the earth's orbital motion, $\alpha =0$
around second of June. In the laboratory frame we have
\cite{JDV99b}
$ {\bf \upsilon}={\bf \upsilon}^{'} \, - \, {\bf \upsilon}_E \,$ 

\section{Event Rates}

Integrating Eq. (\ref {2.18}) we obtain for the total
non-directional rate
\begin{equation}
R =  \bar{R}\, t \, \frac{2 \bar{\rho}}{\rho(0)}
          [1 - h(a,Q_{min})cos{\alpha}] 
\label{3.55}  
\end{equation}
The integration was performed from$u=m_{min}$ to $u=u_{max}$, where
\begin{eqnarray}
u_{min}= \frac{Q_{min}}{Q_0},
u_{max}=min(\frac{y^2_{esc}}{a^2},max(\frac{y_{n} ^2}{a^2})~,~ n=1,2,...,N)
\label{3.30c}  
\end{eqnarray}
Here $y_{esc}=\frac{\upsilon_{esc}}{\upsilon_0}$, with 
$\upsilon_{escape}=625 Km/s$
is the escape velocity from the galaxy.
$Q_{min}$ is the energy transfer cutoff imposed by the detector
and $a =[\sqrt{2} \mu _rb\upsilon _0]^{-1}.  $ Also
$\rho_{n}=d_n/\bar{\rho}
,\bar{\rho}=\sum_{n=1}^N~d_n$ (for each flow +,-).
In the Sikivie model
\cite {SIKIVIE} $2\bar{\rho}/\rho(0)=1.25$. 
$\bar{R}$ is obtained  
\cite {JDV} by neglecting the folding with the LSP velocity and the
momentum transfer dependence, i.e. by
\begin{eqnarray}
\bar{R}& =&\frac{\rho (0)}{m_{\chi}} \frac{m}{Am_N} \sqrt{\langle
v^2\rangle } [\bar{\Sigma}_{S}+ \bar{\Sigma} _{spin} + 
\frac{\langle \upsilon ^2 \rangle}{c^2} \bar{\Sigma} _{V}]
\label{3.39b}  
\end{eqnarray}
and it contains all SUSY parameters except $m_{\chi}$
 The modulation is described in terms of the parameter $h$. 
 The effect of folding
with LSP velocity and the nuclear form factor is taken into account by 
$t$ (see table 1) 

\begin{table}
\caption{The quantities $t$ and $h$ entering the total non-directional
rate in the case of the
target $_{53}I^{127}$ for various LSP masses and $Q_{min}$ in KeV. 
Also shown are the quantities $r^i_j,h^i_j$
 $i=u,d$ and $j=x,y,z,c,s$, entering the directional rate for no energy
cutoff. For definitions see text. 
}
\begin{center}
\renewcommand{\arraystretch}{1.4}
\setlength\tabcolsep{5pt}
\begin{tabular}{|c|c|rrrrrrr|}
\hline
& & & & & & & &     \\
&  & \multicolumn{7}{|c|}{LSP \hspace {.2cm} mass \hspace {.2cm} in GeV}  \\ 
\hline 
& & & & & & & &     \\
Quantity &  $Q_{min}$  & 10  & 30  & 50  & 80 & 100 & 125 & 250   \\
\hline 
& & & & & & & &     \\
t      &0.0&1.451& 1.072& 0.751& 0.477& 0.379& 0.303& 0.173\\
h      &0.0&0.022& 0.023& 0.024& 0.025& 0.026& 0.026& 0.026\\
$r^u_z$&0.0&0.726& 0.737& 0.747& 0.757& 0.760& 0.761& 0.761\\
$r^u_y$&0.0&0.246& 0.231& 0.219& 0.211& 0.209& 0.208& 0.208\\
$r^u_x$&0.0&0.335& 0.351& 0.366& 0.377& 0.380& 0.381& 0.381\\
$h^u_z$&0.0&0.026& 0.027& 0.028& 0.029& 0.029& 0.030& 0.030\\
$h^u_y$&0.0&0.021& 0.021& 0.020& 0.020& 0.019& 0.019& 0.019\\
$h^u_x$&0.0&0.041& 0.044& 0.046& 0.048& 0.048& 0.049& 0.049\\
$h^u_c$&0.0&0.036& 0.038& 0.040& 0.041& 0.042& 0.042& 0.042\\
$h^u_s$&0.0&0.036& 0.024& 0.024& 0.023& 0.023& 0.022& 0.022\\
$r^d_z$&0.0&0.274& 0.263& 0.253& 0.243& 0.240& 0.239& 0.239\\
$r^d_y$&0.0&0.019& 0.011& 0.008& 0.007& 0.007& 0.007& 0.007\\
$r^d_x$&0.0&0.245& 0.243& 0.236& 0.227& 0.225& 0.223& 0.223\\
$h^d_z$&0.0&0.004& 0.004& 0.004& 0.004& 0.004& 0.004& 0.004\\
$h^d_y$&0.0&0.001& 0.000& 0.000& 0.000& 0.000& 0.000& 0.000\\
$h^d_x$&0.0&0.022& 0.021& 0.021& 0.020& 0.020& 0.020& 0.020\\
$h^d_c$&0.0&0.019& 0.018& 0.018& 0.017& 0.017& 0.017& 0.017\\
$h^d_s$&0.0&0.001& 0.001& 0.000& 0.000& 0.000& 0.000& 0.000\\
\hline 
& & & & & & & &     \\
t    &10.0&0.000& 0.226& 0.356& 0.265& 0.224& 0.172& 0.098\\
h    &10.0&0.000& 0.013& 0.023& 0.025& 0.025& 0.026& 0.026\\
\hline 
& & & & & & & &     \\
t    &20.0&0.000& 0.013& 0.126& 0.139& 0.116& 0.095& 0.054\\
h    &20.0&0.000& 0.005& 0.017& 0.024& 0.025& 0.026& 0.026\\
\hline
\end{tabular}
\end{center}
\label{apptab1.1b}
\end{table}

 There are now experiments under way aiming at measuring directional rates
, i.e. the case in which the nucleus is observed in a certain direction.
The rate will depend on the direction of observation, showing a strong
correlation with the direction of both the sun's and the earth's motion. In 
the  favorable 
situation the rate will merely be suppressed by about a factor of $2 \pi$
relative to the non-directional rate. This is due to the fact that one 
does not now integrate over the
azimuthal angle of the nuclear recoiling momentum. 

 We need  distinguish the following cases: a) $\hat{e}$ has a
component in the sun's direction of
motion, i.e. $0<\theta < \pi /2$, labeled by i=u (up). b) Detection
in the opposite direction, $\pi /2 <\theta < \pi $, labeled by 
i=d (down).  
Thus we find :

1. In the first quadrant (azimuthal angle $0 \leq \phi \leq \pi/2)$.
\begin{eqnarray}
R^i_{dir} & = &\bar{R} \frac{2 \bar{\rho}}{\rho(0)}
    \frac{t}{2 \pi} [(r^i_z  - \cos \alpha~ h^i_1) {\bf e}_z.{\bf e}  
\nonumber \\ 
&+& (r^i_y +cos \alpha h^i_2 +\frac{h^i_c }{2}(|cos\alpha|+cos\alpha))
     |{\bf e}_y.{\bf e} | 
\nonumber \\ 
&+& (r^i_x -sin \alpha h^i_3 +\frac{h^i_s }{2}(|sin\alpha|-sin\alpha))
     |{\bf e}_x.{\bf e} | ]
\label{3.56}  
\end{eqnarray}
2. In the second quadrant (azimuthal angle $\pi/2 \leq \phi \leq \pi)$
\begin{eqnarray}
R^i_{dir} & = &\bar{R} \frac{2 \bar{\rho}}{\rho(0)}
    \frac{t}{2 \pi}  [(r^i_z  - \cos \alpha~ h^i_1) {\bf e}_z.{\bf e}  
\nonumber \\ 
&+& (r^i_y +cos \alpha h^i_2 (u)+\frac{h^i_c }{2}(|cos\alpha|-cos\alpha))
     |{\bf e}_y.{\bf e} | 
\nonumber \\ 
&+& (r^i_x +sin \alpha h^i_3 +\frac{h^i_s }{2}(|sin\alpha|+sin\alpha))
     |{\bf e}_x.{\bf e} | ]
\label{3.57}  
\end{eqnarray}
3. In the third quadrant (azimuthal angle $\pi \leq \phi \leq 3 \pi/2)$.
\begin{eqnarray}
R^i_{dir} & = &\bar{R} \frac{2 \bar{\rho}}{\rho(0)}
    \frac{t}{2 \pi}  [(r^i_z  - \cos \alpha~ h^i_1) {\bf e}_z.{\bf e}  
\nonumber \\ 
&+& (r^i_y -cos \alpha h^i_2 (u)+\frac{h^i_c (u)}{2}(|cos\alpha|-cos\alpha))
     |{\bf e}_y.{\bf e} | 
\nonumber \\ 
&+& (r^i_x +sin \alpha H^i_3 +\frac{h^i_s }{2}(|sin\alpha|+sin\alpha))
     |{\bf e}_x.{\bf e} | ]
\label{3.58}  
\end{eqnarray}
4. In the fourth quadrant (azimuthal angle $3 \pi/2 \leq \phi \leq 2 \pi)$
\begin{eqnarray}
R^i_{dir} & = &\bar{R} \frac{2 \bar{\rho}}{\rho(0)}
    \frac{t}{2 \pi}  [(r^i_z  - \cos \alpha~ h^i_1) {\bf e}_z.{\bf e}  
\nonumber \\ 
&+& (r^i_y -cos \alpha h^i_2 +\frac{h^i_c }{2}(|cos\alpha|-cos\alpha))
     |{\bf e}_y.{\bf e} | 
\nonumber \\ 
&+& (r^i_x -sin \alpha h^i_3 +\frac{h^i_s }{2}(|sin\alpha|-sin\alpha))
     |{\bf e}_x.{\bf e} | ]
\label{3.59 }  
\end{eqnarray}

\section{Conclusions}
 We have calculated the parameters describing
characteristic signatures needed 
to reduce the formidable backgrounds
in the direct detection of SUSY dark matter,
 such as : a) The modulation effect, 
correlating the rates with the motion of the Earth  and b)
 The directional  rates, correlated with both with the velocity of the sun
and that of the Earth (see table 1).

 We have focused on the LSP density and velocity spectrum 
obtained from a recently proposed non-isothermal model, involving caustic
rings 
\cite{SIKIVIE}. Our results for isothermal models have appeared
elsewhere 
\cite{JDV99,JDV99b}. 

 The quantities  $t$ and $h$ are given in table 1. 
We see that the maximum in this model does not  
occur around June 2nd, but about six months later. The difference
between the maximum and the minimum is about $4\%$, i.e. smaller
than that predicted by the asymmetric isothermal models 
\cite{JDV99,JDV99b}. 

 For the directional experiments we found that 
the biggest rates are obtained close to the direction of the sun's motion.
They are suppressed compared to the usual 
non-directional rates by the factor 
$f_{red}=\kappa/(2 \pi)$, $\kappa=u^i_z$. We find $\kappa \simeq 0.7$, 
$i=up$ (observation in the sun's direction of motion) while
$\kappa\simeq 0.3$, $i=down$ ( in the opposite direction).
The modulation is a bit larger than in the non-directional case. The
largest difference between the maximum and the minimum, 
 $8\%$, occurs  not the sun's direction of motion, but in the x-direction
 (galactocentric direction).

 In the case of the isothermal models the reduction factor 
along the sun's direction of motion
is now given $f_{red}=t_0/(4 \pi~t)=\kappa/(2 \pi)$. 
Using the values of $t_0$ obtained previously
\cite{JDV99b},
we find
that $\kappa$ is around 0.6 for the symmetric case  and around 0.7 for maximum
asymmetry ($\lambda=1.0$). 
The modulation of the directional rate depends on the direction of 
observation. It is generally larger and 
increases with the asymmetry parameter $\lambda$. 
 For $Q_{min}=0$ it can reach values up 
to $23\%$. Values up to $35\%$ are possible for large
 $Q_{min}$, but at the expense of the total number of counts
\cite{JDV99b}.

 Finally in  all cases $t$
deviates from unity for large reduced mass.
Thus when extracting the LSP-nucleon cross section from
the data one must divide by $t$. 

{\it Acknowledgments:} The author happily 
acknowledges partial support of this work by
TMR No ERB FMAX-CT96-0090 of the European Union. 

\end{document}